\documentclass{emulateapj}
\usepackage{amsmath} 
\usepackage{apjfonts} 
\usepackage{amssymb} 
\usepackage{xspace}
\usepackage{hyperref} 
\usepackage{natbib} 
\usepackage{multirow}
\usepackage{graphicx}
\usepackage{epstopdf}
\usepackage{epsf}
\usepackage{subfigure}
\usepackage[usenames,dvipsnames]{xcolor}

\def\gsim{\gtrsim}
\def\lsim{\lesssim}    

\begin{document} 
\shorttitle{Temperature Evolution in Cluster Outskirts}
\shortauthors{Avestruz, Nagai, Lau}
\submitted{The Astrophysical Journal} 
\slugcomment{The Astrophysical Journal, submitted}

\title{Stirred, not Clumped: Evolution of Temperature Profiles in the
  Outskirts of Galaxy Clusters}

\author{Camille Avestruz\altaffilmark{1-5}\thanks{E-mail:
    avestruz@uchicago.edu}, Daisuke Nagai\altaffilmark{1,2}, and Erwin
  T.\ Lau\altaffilmark{1,2} }

\affil{
$^1${Department of Physics, Yale University, New Haven, CT 06520, U.S.A.};\\
$^2${Yale Center for Astronomy \& Astrophysics, Yale University, New Haven, CT 06520, U.S.A.} \\
$^3${Enrico Fermi Institute, The University of Chicago, Chicago, IL 60637 U.S.A.}\\
$^4${Kavli Institute for Cosmological Physics, The University of Chicago, Chicago, IL 60637 U.S.A.}\\
$^5${Department of Astronomy \& Astrophysics, The University of Chicago, Chicago, IL 60637 U.S.A.};
  \href{mailto:avestruz@uchicago.edu}{avestruz@uchicago.edu}\\
}

  
\begin{abstract} 
  Recent statistical X-ray measurements of the intracluster medium
  (ICM) indicate that gas temperature profiles in the outskirts of
  galaxy clusters deviate from self-similar evolution.  Using a
  mass-limited sample of galaxy clusters from cosmological
  hydrodynamical simulations, we show that the departure from
  self-similarity can be explained by non-thermal gas motions driven
  by mergers and accretion. Contrary to previous claims, gaseous
  substructures only play a minor role in the temperature evolution in
  cluster outskirts.  A careful choice of halo overdensity definition
  in self-similar scaling mitigates these departures. Our work
  highlights the importance of non-thermal gas motions in ICM
  evolution and the use of galaxy clusters as cosmological probes.
\end{abstract}

\section{Introduction}

The outskirts of galaxy clusters mark the transition from the cosmic
web to the intracluster medium (ICM), where cosmic filaments feed into
the gravitational potential well of the galaxy cluster.  Measurements
of gas properties in cluster outskirts directly probe the formation of
these structures and physical mechanisms that affect the ICM.  A
thorough understanding of these mechanisms at all radii will further
enable cluster cosmology.

The use of clusters as cosmological probes hinges on our ability to
tightly constrain the evolution of the galaxy cluster mass function,
which is sensitive to cosmological parameters \citep[see][for a
  review]{allen_etal11}.  We can exploit self-similar properties of
galaxy clusters to establish relationships between directly observable
properties and the galaxy cluster mass \citep[see][for a
  review]{voit_05,kravtsovandborgani_12}.

The ICM radial profiles of self-similar galaxy clusters resemble one
another when rescaled with respect to mass and redshift dependent
quantities.  When radially integrated, profile quantities can serve as
a proxy of cluster mass.  The self-similar model of galaxy clusters is
based on simplifying assumptions. In reality these assumptions are
broken in our universe's cosmology and through non-gravitational
baryonic processes, such as radiative cooling and star formation.  The
only way to directly probe and exploit self-similar relations is to
self-consistently model galaxy cluster formation and calibrate
observable-mass relations used by observations. To mitigate
complications from baryonic physics that dominate the cluster cores,
observations have pushed to cluster outskirts to potentially recapture
self-similar behavior.  Evidence, however, points to the contrary.

First, X-ray measurements show that the ICM entropy profiles break
from the self-similar power-law scaling with radius at large radii
\citep[e.g.][]{george_etal09,bautz_etal09,reiprich_etal09,hoshino_etal10,
  kawaharada_etal10,akamatsu_etal11,okabe_etal14} contradicting
theoretical predictions \citep{tozziandnorman_01, voit_etal05}.  Next,
the observational analysis of \citet{mcdonald_etal14} presents
evidence of a non-self-similar evolution of ICM properties at radii
larger than ${R}_{500c}$.  With {\em Chandra} data of 80 galaxy
clusters selected from the {\em South Pole Telescope} (SPT) survey,
they found that the outskirts of high redshift clusters have cooler
scaled temperatures than low redshift clusters.  The origin of the
observed temperature evolution is well not understood.

To address the first finding, detailed X-ray observations with {\em
  Suzaku} \citep{simionescu_etal11,walker_etal13,urban_etal14} and
{\em ROSAT} \citep{eckert_etal12,eckert_etal13} provided evidence that
density inhomogeneities in the ICM likely break self-similar entropy
profiles \citep[but see][for a different
  explanation]{okabe_etal14}. In support of observational evidence,
cosmological simulations showed that gas in cluster outskirts are
highly clumpy \citep{nagaiandlau_11,roncarelli_etal13, vazza_etal13}.
This clumpiness boosts the X-ray surface brightness, leading to
overestimates in gas densities and underestimates in entropy
\citep{avestruz_etal14}.  Such ICM inhomogeneities are associated with
overdense infalling substructures such as subhalos and penetrating
filaments \citep{battaglia_etal15, lau_etal15}.

To extend clumping effects to the second finding,
\citet{mcdonald_etal14} proposed that the presence of more accreting
group size halos at higher redshifts drives the evolution of the
scaled outskirt temperatures, an effect called ``superclumping''.
This interpretation is based on the fact that mass accretion rates and
merger frequencies increase with redshift
\citep[e.g.,][]{wechsler_etal02,fakhouriandma_09,tillson_etal11}, and
that group size halos substructures have lower virial temperatures
than host cluster ICM.  At high redshift, these substructures are
usually too dim for X-ray observations to identify and mask out.

On the other hand, a significant fraction of outskirt gas does not
reside in these dense substructures.  Rather, much of the outskirt gas
exists in the form of low density, diffuse gas
\citep[e.g.,][]{zhuravleva_etal13}.  In the outskirts, accretion
shocks convert kinetic energy of infalling gas to thermal energy.
However, not all of the kinetic energy is thermalized by the shocks.
The residual gas motions slowly thermalize through turbulent
dissipation at the dynamical timescale \citep{shiandkomatsu_14}.  As a
result, clusters that experience more recent gas accretion have less
time to thermalize and tend to have a higher fraction of energy in the
form of non-thermal gas motions
\citep{lau_etal09,vazza_etal11,nelson_etal12, shi_etal15}.  Therefore,
the ICM temperature evolution can also depend on the rate at which
these gas motions thermalize.

The aim of this work is to understand the physical mechanisms that
contribute to the observed non-self-similar evolution of temperature
profiles in cluster outskirts from the X-ray observations of the {\em
  Chandra}-SPT cluster sample by \citet{mcdonald_etal14}.  We use a
mass-limited sample of galaxy clusters from a cosmological
hydrodynamical simulation.  Our sample is comparable in mass and
redshift range to the observed cluster sample.  We demonstrate that
non-thermal gas motions in clusters is the dominant contributor to the
observed temperature evolution in cluster outskirts.  Substructure
evolution plays a subdominant role, in contrast to the original
interpretation by \citet{mcdonald_etal14}.

Our paper is organized as follows.  We overview the notions of
self-similarity in Section~\ref{sec:self-similarity} and define ICM
quantities of interest in Section~\ref{sec:pressure}. In
Section~\ref{sec:methods} we briefly describe the simulation and the
radial profile averaging.  We present our results in
Section~\ref{sec:results}, and our summary and discussions in
Section~\ref{sec:conclusions}.

\section{Theoretical Framework}

\subsection{Self-similar model}
\label{sec:self-similarity}

The standard self-similar model by \citet{kaiser_86} describes the
properties of galaxy clusters based on their mass and redshifts.  This
Kaiser model is based on several simplifying assumptions about the
formation of galaxy clusters.  First, galaxy clusters form from
scale-free gravitational collapse of the initial density perturbations
in an $\Omega_m=1$ universe.  Second, the amplitude of initial density
fluctuation is scale-free; i.e., the matter power spectrum is a power
law with $P(k)\propto k^n$.  Finally, there are no additional physical
processes that introduce any scale dependence.  The scale-free setup
of this problem defines a self-similar model where halo properties
depend only on the slope and normalization of the initial density
field at collapse.

Assuming further that the cluster gas is spherically symmetric and is
in hydrostatic equilibrium with the cluster's gravitational potential
well, we can define a characteristic temperature $T_\Delta$, which
relates to the cluster mass as,
\begin{eqnarray}\label{eqn:TM_relation}
  {T_\Delta}\propto
  \frac{GM_\Delta}{R_\Delta}\propto(\Delta\rho_r)^{1/3}M_\Delta^{2/3}.
\end{eqnarray}
Here, $M_\Delta$ is the mass enclosed within a sphere of radius,
$R_\Delta$, both defined with respect to some reference density
$\rho_{\rm r}$ such that,
\begin{eqnarray}\label{eqn:mass_def}
  M_\Delta = \frac{4\pi}{3}R_\Delta^3 \Delta \rho_{\rm r}(z) ,
\end{eqnarray}
and $\Delta$ is the mean overdensity contrast with respect to a
reference density, $\rho_r(z)$, at a given redshift $z$. The 
characteristic temperature is then a function of cluster mass and redshift
$T_\Delta \equiv T_\Delta (M_\Delta,z)$. 

The temperature profile can be then scaled with respect to this
characteristic temperature as,
\begin{eqnarray}\label{eqn:Tparam}
\tilde{T}\left(r/{R_\Delta}\right)&\equiv& \frac{T\left(r/R_\Delta
  \right)}{T_\Delta(M_\Delta,z)}.
\end{eqnarray}
If galaxy clusters were perfectly self-similar under this scaling, the
shape and normalization of $\tilde{T}(r/R_\Delta)$ would be
independent of the cluster mass and redshift.  

It is important to note that a mass-limited sample of real galaxy
clusters do not satisfy {\it any} of the assumptions of the Kaiser
model, but have nonetheless empirically exhibited near self-similar
behavior.  We therefore explore physical effects that lead to observed
deviations from self-similar scaling in a similarly representative
sample.

Note that the self-similar model depends on which overdensity is used
to define cluster mass and radius in Equation~(\ref{eqn:mass_def}).  A
commonly used density contrast in cluster measurements is $\Delta =
500c$, where $c$ denotes the overdensity defined with respect to the
critical density of the universe.  We additionally consider several
different characteristic overdensities $\Delta=200c, 500c, 1600c$ and
$\Delta = 200m, 500m, 1600m$, where $m$ denotes the overdensity
defined with respect to the mean mass density of the universe. Note
that these overdensities are redshift dependent; e.g., $\Delta=1600m$
approximately corresponds to $\Delta=500c$ at $z=0$.

\begin{figure*}[t]
\begin{center}
\includegraphics[scale=0.55]{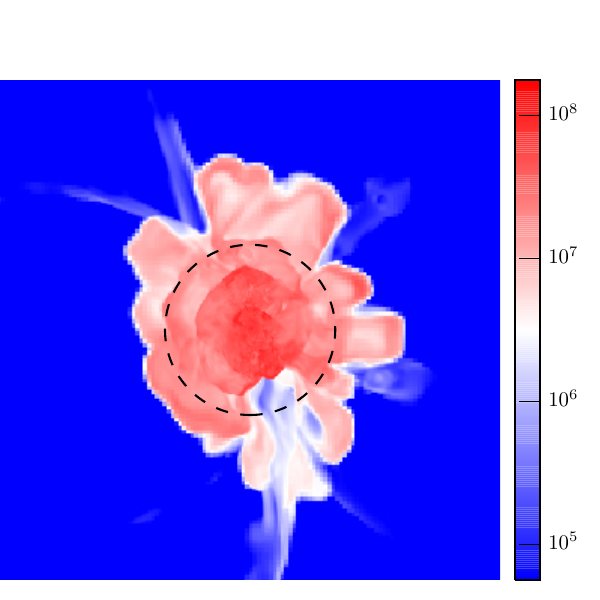}
\includegraphics[scale=0.55]{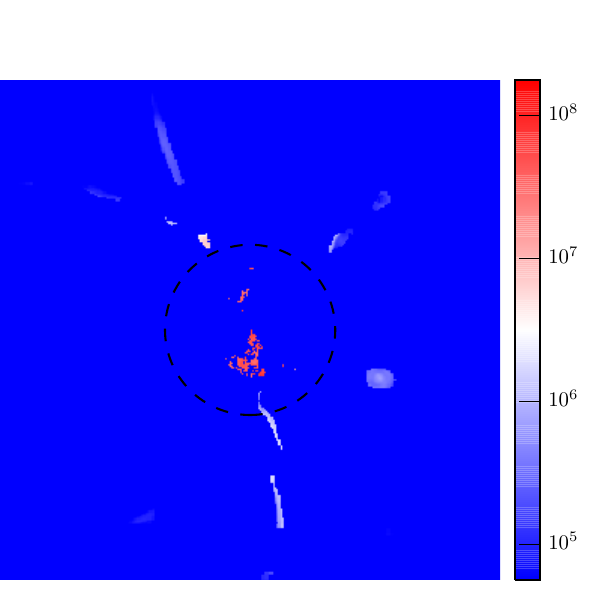}
\includegraphics[scale=0.55]{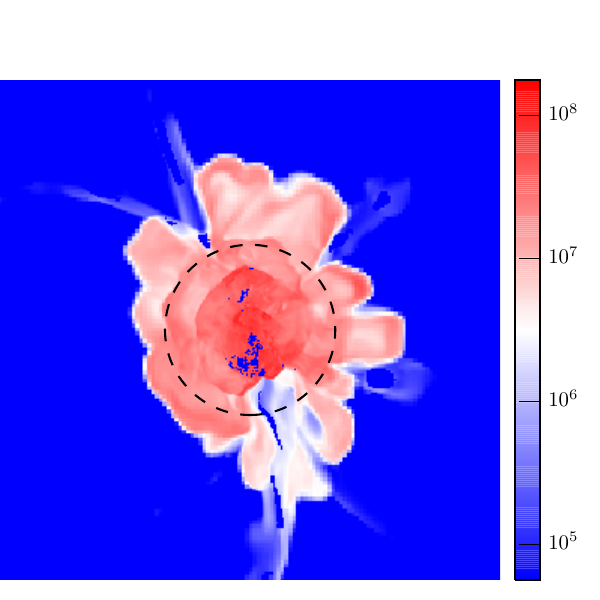}
\caption
{Projected mass-weighted temperature $T_\text{mw}$ maps of one of the
  $z=0$ clusters selected from the sample. From {\em left} to {\em
    right}: map of the total gas, substructures (clump and filaments),
  and ``bulk'' gas (i.e., without substructures). The dimension for
  each panel is $15.6\,h^{-1}{\rm Mpc}\times15.6\,h^{-1}{\rm Mpc}$,
  with depth of $1.9\,h^{-1}{\rm Mpc}$. The circle in dashed line
  shows $3R_{500c} = 2.6\,h^{-1}{\rm Mpc}$ of the
  cluster. The color bars show the temperature scales in Kelvin. }
\label{fig:cluster_map}
\end{center}
\end{figure*}

\subsection{Thermal, Non-thermal and Total Energy Contents of the ICM}\label{sec:pressure}

Energy content of the ICM consists of thermal and non-thermal
components, where the latter arises primarily from random gas motions
generated by mergers and accretion events.  We can define the
``total'' temperature as the sum of specific kinetic and internal
energies of the gas, as it incorporates both thermal and non-thermal
motions of the gas,
\begin{eqnarray}
T_{\text{tot}}&\equiv&T_{\text{mw}}+T_{\text{nt}},
\end{eqnarray}
where $T_{\text{mw}}$ is the mass-weighted temperature and
$T_{\text{nt}}$ is the ``non-thermal temperature'' defined as,
\begin{eqnarray}\label{eqn:tnt}
k_bT_\text{nt}&\equiv&\frac{1}{3}\mu m_p \langle
v^2_{\text{gas}}\rangle_{\rm mw},
\end{eqnarray}
where $k_b$ is the Boltzmann constant, $\mu=0.59$ is the mean
molecular weight of the ionized ICM, $m_p$ is the proton mass, and
$\sqrt{\langle v^2_{\text{gas}}\rangle_{\rm mw}}$ is the 3-dimensional
mass-weighted root-mean-square velocity of the gas.  Physically, this
``non-thermal temperature'' $k_bT_\text{nt}$ represents the specific
kinetic energy associated with gas motions in the ICM.

The total temperature for a virialized gas is analogous to the
velocity dispersion of a halo, which is in local Jeans equilibrium
with the gravitational potential of the halo \citep{diemer_etal13}.
The circular velocity profile, a proxy for the potential, evolves
self-similarly when scaled with respect to the critical density of the
universe \citep{diemerandkravtsov_14, lau_etal15}.  This follows from
the fact that gravitational potential wells of galaxy clusters are set
early in their history \citep[e.g.,][]{vandenbosch_etal14} and evolve
slowly.  Therefore, the total temperature profile within near
virialized regions should exhibit the same scaling as the circular
velocity profile.  Note that the thermal temperature would only
exhibit this scaling if the cluster were in perfect hydrostatic
equilibrium.

\section{Methodology}
\label{sec:methods}

\subsection{Cosmological Simulation}

We use galaxy clusters extracted from the {\em Omega500} simulation
\citep{nelson_etal14}.  The {\em Omega500} simulation is a
cosmological hydrodynamical simulation performed with the Adaptive
Refinement Tree code \citep{kravtsov_99, kravtsov_etal02,
  rudd_etal08}. The simulation box has a comoving length of
500~$h^{-1}\, {\rm Mpc}$, resolved using a uniform $512^3$ root grid
and 8 levels of mesh refinement, with maximum comoving spatial
resolution of 3.8~$h^{-1}\, {\rm kpc}$.  
    
We analyze a mass-limited sample of 65 galaxy clusters with $M_{\rm
  500c}\geq 3\times10^{14}h^{-1}M_{\odot}$ at $z=0.0$ and their
progenitors at $z=0.3$, $0.5$, and $0.7$ to match the {\em
  Chandra}-SPT sample of \citet{mcdonald_etal14}.  We then measure the
average evolution of thermodynamic profiles in our sample.

Initial cluster identification uses a spherical overdensity halo
finder described in \citet{nelson_etal14}.  The final cluster sample
is from a re-simulated box with higher resolution dark matter
particles in regions of the identified clusters.  This ``zoom-in''
technique results in an effective mass resolution of $2048^{3}$,
corresponding to a dark matter particle mass of $9\times10^{8}\,
h^{-1} M_{\odot}$ inside a spherical region with cluster-centric
radius of three times the virial radius for each cluster.  Further
details of the simulation can be found in \citet{nelson_etal14}.

Since cluster core physics are not expected to significantly affect
cluster outskirts, we present our main results using the {\em
  Omega500} simulation with non-radiative (NR) gas physics.  In order
to assess the effects of baryonic physics, we also analyze the outputs
of the {\em Omega500} re-simulation with radiative cooling, star
formation, and supernova feedback based on the same sub-grid model of
galaxy formation described in \citet{nagai_etal07a}.

\subsection{Averaged Radial Profiles}\label{sec:meth:avg_rad}

We compute average radial profiles by dividing the gas volume of a
galaxy cluster halo into 99 concentric spherical shell bins centered
around the minimum of the gravitational potential.  Throughout this
work, we use the equally spaced logarithmic spacing in the comoving
radial distance from $10\,h^{-1}$~kpc to $10\,h^{-1}$~Mpc.

We compute average temperature profile of each simulated galaxy
cluster as,
\begin{equation}
T_{w} (r_i)=\frac{\sum_j w_{ij} T_{ij}\Delta V_j}{\sum_j w_{ij}\Delta V_{ij}},
\end{equation}
where $\Delta V_{ij}$ is the volume occupied by the hydro cell $j$ in
the radial bin $i$, $T_{ij}$ is the temperature of the gas cell, and
$w_{ij}$ is the ``weight'' for the averaging.

We adopt two different weighting schemes. First,  we compute mass-weighted 
temperature profile $T_{\text{mw}}$, with the weight set to the gas mass density 
of the hydro cell $w_{ij} = \rho_{g,ij}$. Physically, the mass-weighted temperature 
corresponds to the specific internal energy of the cluster gas. 

Second, we compute the spectral-temperature $T_\text{sp}$, which is
the average temperature using X-ray emission as weights: 
\begin{equation}\label{eqn:emission}
w_{ij} =  \rho_{g,ij} ^2 \Lambda_{\rm eff}(T_{ij},Z_{ij}=0.3Z_\odot) 
\end{equation}
where $\Lambda_{\rm eff}$ is the effective cooling function in the
$0.5-2.0$~keV energy band using the MEKAL \citep{liedahl_etal95}
plasma code, weighted by the effective area of the {\em ACIS-I} CCD on
{\em Chandra} X-ray telescope.  Since the X-ray emission in the
$0.5-2.0$~keV energy range is not very sensitive to the adopted
metallicity, we assume constant abundance of $Z=0.3$ solar throughout.

Finally, using the self-similar scaling defined by
Equation~(\ref{eqn:Tparam}), we normalize each galaxy cluster profile,
\begin{eqnarray}
\tilde{T}(r/R_{500c})&=&\frac{T(r/R_{500c})}{T_{500c}},
\end{eqnarray}
where $T_{500c}\equiv GM_{500c}/(2R_{500c})$.  We then calculate the
average $\tilde{T}$ in each redshift in order to assess the departure
of the {\em average normalized temperature profile} from self-similar
evolution.

\begin{figure}
  \centering 
  \mbox{
  \includegraphics[width=3.5in]{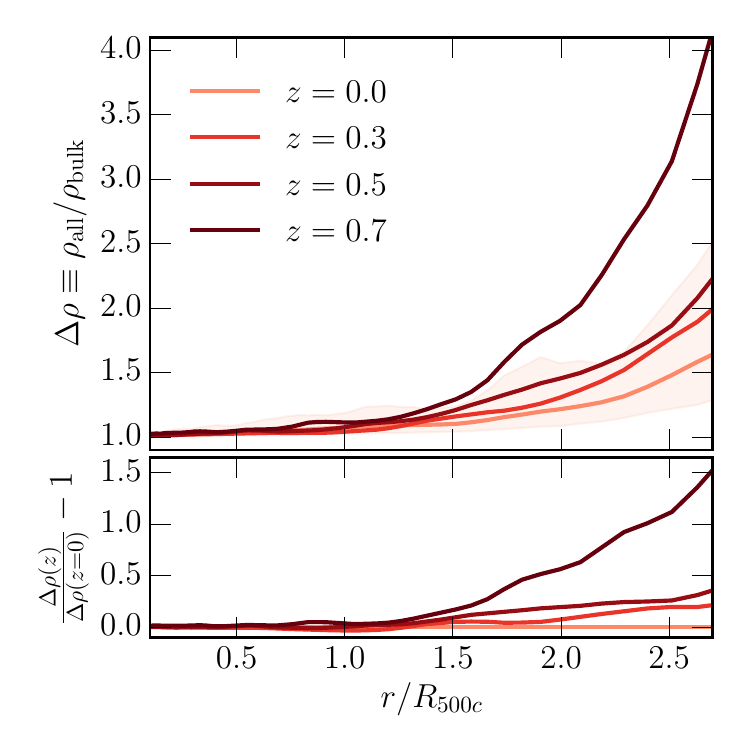}
  }
  \vspace{-5mm}
  \caption{Top panel: Ratio of the average density profile of all gas
    to that of the bulk component (without substructures), $\Delta
    \rho \equiv \rho_{\rm all}/\rho_{\rm bulk}$, at different
    redshifts as a function of the cluster-centric radius for
    simulated clusters at $z=0.0, 0.3, 0.5, 0.7$.  The shaded region
    corresponds to the $1\sigma$ spread about the mean profile at
    $z=0$. The bottom panel shows the fractional difference between
    the profile at each redshift to the profile at $z=0$.}
  \label{fig:clumping_redshift}
\end{figure}

\subsection{Decomposition of Diffuse and Clumpy Components}\label{sec:meth:clumping}

To assess the effects of gas clumping on the X-ray emissions, we
decompose the ICM into the diffuse and clumpy components using the
method described in \citet{zhuravleva_etal13}.  The probability
distribution function (PDF) of the density in each shell follows a
log-normal distribution with a high density tail.  We exclude small
scale gas clumps, infalling subhalos, and penetrating filaments by
removing gas that have density higher than $2\sigma$ from the median
of the density PDF in each radial bin.

We denote the profiles calculated with or without denser substructures
using the subscripts, ``all'' or ``bulk'', respectively.  The
mass-weighted temperature profile with (or without) substructures is
labeled by $T_{\text{mw,all}}$ (or $T_{\text{mw,bulk}}$).
Figure~\ref{fig:cluster_map} illustrates the mass-weighted temperature
maps for one of the simulated clusters including all gas,
substructures, and bulk diffuse components \citep[see Figure~1 in][for
  the corresponding projected gas density maps]{lau_etal15}.

\section{Results}
\label{sec:results}

In this work, we investigate physical mechanisms that contribute to
the evolution of scaled thermal temperature profiles in the outskirts
of galaxy clusters. These mechanisms include the evolution of
overdense gas substructures, non-thermal gas motions, growth of the
cluster halo with respect to the defined reference overdensity,
non-equilibrium physics, and baryonic cooling and star formation.
With a mass-limited simulated cluster sample, we assess the relative
importance of factors that influence temperature profiles in galaxy
cluster outskirts.

\subsection{Evolution of Substructures}
\label{sec:substructures}

We first examine the contribution of substructures on the evolution of
the gas density profiles.  The top panel of
Figure~\ref{fig:clumping_redshift} shows the ratio of the average
density profile of all gas to that of the bulk component (without
substructures), $\Delta \rho \equiv \rho_{\rm all}/\rho_{\rm bulk}$,
at different redshifts. We normalize the radial range of all profiles
using $R_{500c}$ of each cluster before computing the average profiles
of the cluster sample at each redshift. The lower panel shows the
fractional deviation of the gas density ratio at high redshifts
relative to $z=0$.

At all redshifts, we find that the bulk component comprises more than
$90\%$ of the gas at $r\lesssim R_{500c}$, while the contribution of
substructures increases with radius at $r\gtrsim R_{500c}$. The
contribution of gaseous subhalos and filaments are more prominent in
the outskirts of galaxy clusters. We also find that the contribution
of dense substructures increases with redshift. For example, the
substructures comprise $\approx50\%$ of the total gas density at
$r/R_{500c}=2$ at $z=0.7$, while their contribution is only $\approx
20\%$ at $z=0$. High redshift clusters contain an enhanced level of
substructures compared to the low redshift counterparts.

\subsection{Evolution of ICM Temperatures}
\label{sec:non-thermal}

\begin{figure}
  \centering 
  \mbox{
  \subfigure{\includegraphics[width=3.5in]{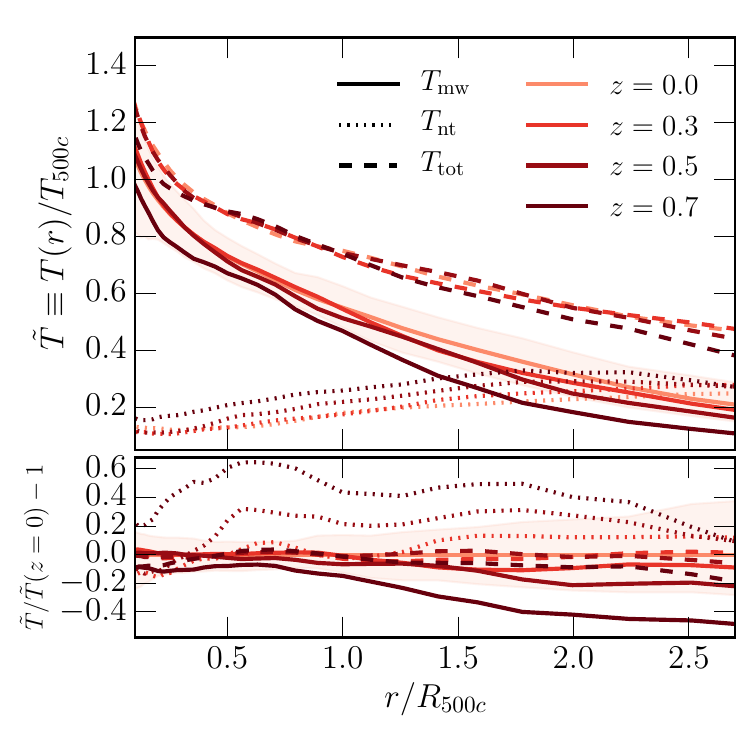}}
  }
  \vspace{-5mm}
  \caption{Top panel: Solid lines correspond to the average normalized
    mass weighted gas temperature profile, $\tilde{T}_{\rm mw}$, at
    $z=0.0, 0.3, 0.5, 0.7$.  The shaded region shows the $1\sigma$
    spread for the $z=0$ sample.  The dotted lines correspond to the
    average normalized non-thermal temperature, $\tilde{T}_{\rm nt}$,
    defined in Equation~\ref{eqn:tnt}.  The dashed lines correspond to
    the sum of the two temperatures at each redshift,
    $\tilde{T}_{\text{tot}}=\tilde{T}_{\text{mw}}+\tilde{T}_{\text{nt}}$. Bottom
    panel: Solid (dashed or dotted) lines show the difference in the
    ratios for between each thermal (total on non-thermal) average
    temperature profile to the $z=0$ average profile.  The normalized
    $\tilde{T}_{\text{mw}}$ profiles have a clear systematic evolution
    outside of $r/R_{500c}\gsim 0.5$ that breaks self-similarity.
    High redshift clusters have cooler outskirts.  The trend is
    reversed for $\tilde{T}_{\text{nt}}$.  The total temperature,
    however, maintains a self-similar behavior at all radii with no
    indication of a systematic evolution.  The thermal temperature
    evolution is due to the fact that clusters at higher redshifts are
    dynamically younger with gas that has not fully
    thermalized.}\label{fig:Tmw_nt_tot}
\end{figure}
\begin{figure*}[t]
  \centering \mbox{
    \subfigure{\includegraphics[width=3.5in]{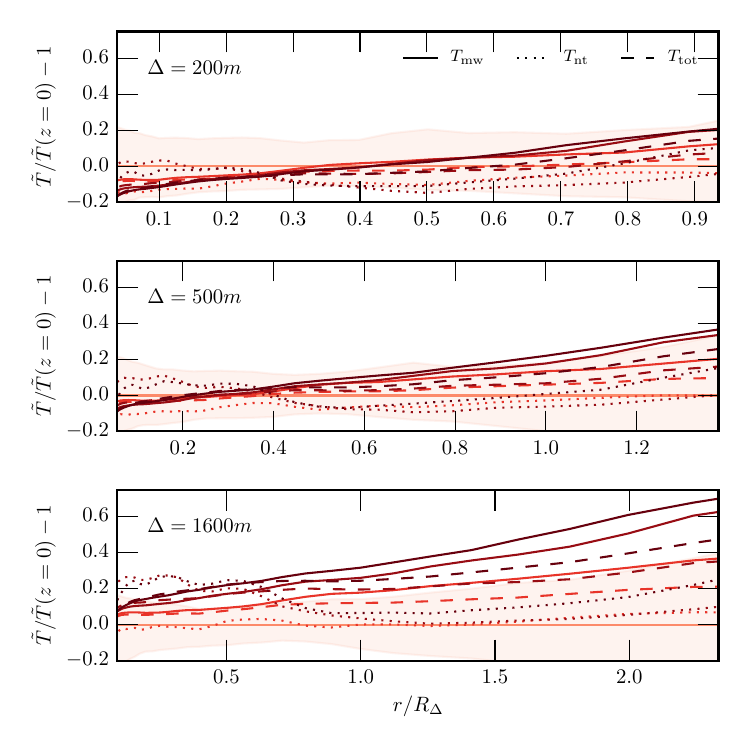}}\quad
    \subfigure{\includegraphics[width=3.5in]{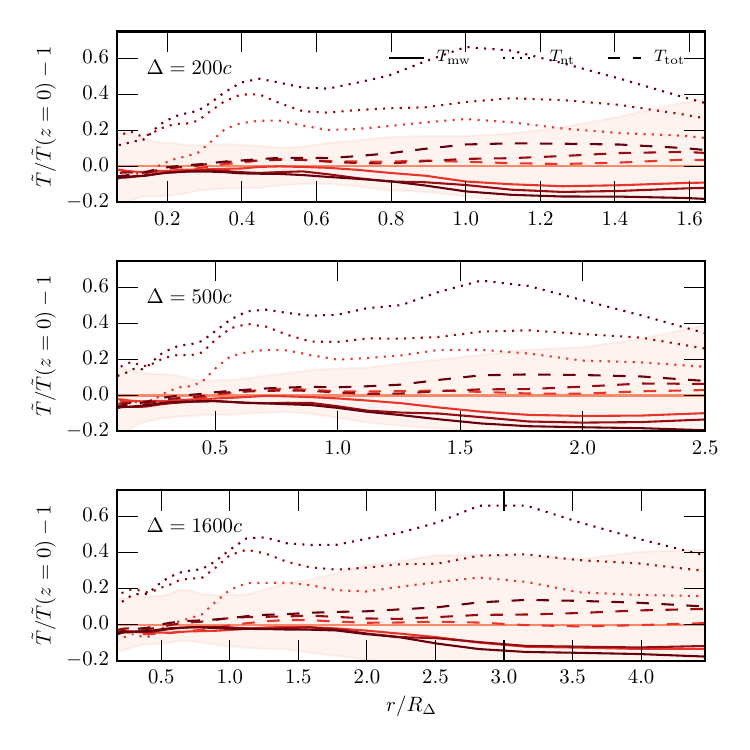}}
  }
  \vspace{-5mm}
  \caption{Same as the lower panel in Figure~\ref{fig:Tmw_nt_tot} but
    for different overdensities.  From top to bottom: $\Delta = 200$,
    $500$, and $1600$.  The left column corresponds to overdensities
    with respect to the mean density of the universe, denoted with
    $m$, and the right with respect to the critical density of the
    universe, denoted with $c$.  The radial scales are chosen such
    that each overdensity probes approximately the same radial range
    $0.1\leq r/R_{500c} \leq 2.7$ at $z=0$.  The scaling with respect
    to $\Delta=200m$ best captures the evolution in thermalization
    state of the gas in the outskirts; the mass-weighted temperature
    exhibits the most self-similar evolution with this choice of
    $\Delta$.}\label{fig:Tmw_nt_tot_overdensities}
\end{figure*}

Next, we examine the role of the non-thermal gas motions on the
evolution of temperature profiles.  The solid lines in the top panel
of Figure~\ref{fig:Tmw_nt_tot} show the average radial profile of the
normalized mass-weighted temperature, $\tilde{T}_\text{mw}$, at
$z=0.0, 0.3, 0.5$, and $0.7$.  The solid lines in the bottom panel of
Figure~\ref{fig:Tmw_nt_tot} show the difference between
$\tilde{T}_\text{mw}$ at each redshift to that of $z=0$: $\Delta
\tilde{T}_\text{mw} \equiv
\tilde{T}_\text{mw}(z)/\tilde{T}_\text{mw}(z=0)-1$.  Note that $\Delta
\tilde{T}_\text{mw} \approx 0$ would indicate that the cluster sample
exhibits self-similar evolution with this scaling.

The magnitude of $\Delta \tilde{T}_\text{mw}$ systematically increases
with redshift in cluster outskirts.  For example, $\Delta
\tilde{T}_\text{mw}$ at $r/R_{500c} \gtrsim 1.5$ is $\sim 30\%$
between $z=0.0$ to $z=0.7$, and $\sim 10\%$ between $z=0.0$ to
$z=0.5$. These evolutionary trends are consistent with the results of
the recent {\em Chandra} measurements \citep{mcdonald_etal14}.

The dotted lines in Figure~\ref{fig:Tmw_nt_tot} show the average
normalized profiles of the non-thermal temperature,
$\tilde{T}_\text{nt}$, at each redshift.  At all redshifts,
$\tilde{T}_\text{nt}$ increases with radius and exceeds
$\tilde{T}_\text{mw}$ at $2 \lesssim r/R_{500c} \lesssim 4$.  The
crossover occurs at smaller cluster-centric radii for higher redshift
clusters, when the potential well also extends to a smaller radius.
This result is consistent with a physical picture that gas motions
generated by mergers and accretion events are gradually converted into
the kinetic energy through shocks and turbulent dissipation at smaller
radii. Since the timescale of turbulent dissipation is shorter in the
dense, inner regions of galaxy clusters \citep{shiandkomatsu_14}, a
larger fraction of the merger or accretion driven gas motions have
thermalized \citep{yu_etal15}.  The timescale difference leads to
monotonically increasing (decreasing) thermal (non-thermal)
temperature profiles.

Furthermore, we find a significant redshift evolution in
$\tilde{T}_\text{nt}$.  At $r/R_{500c}\approx1.5$,
$\tilde{T}_{\text{nt}}$ increases by $40\%$ between $z=0.0$ and
$z=0.7$. High redshift clusters have a higher level of the normalized
non-thermal temperature than low redshift clusters, because high
redshift clusters are dynamically younger with more active gas
accretion events \citep{nelson_etal14b}.

The dashed lines in the top panel of Figure~\ref{fig:Tmw_nt_tot} show
the average total temperature profiles in each redshift bin, which
exhibit a remarkable degree of self-similar evolution at large radii,
unlike the non-thermal and thermal temperature profiles.

We conclude that the evolution in the scaled thermal temperature
profile is driven primarily by the thermalization process of the
cluster gas where merger and accretion induced gas motions gradually
convert into the thermal energy component of the ICM.

\subsection{Dependence on Halo Overdensity Definition}
\label{sec:halo_def}

In this section, we highlight that the departure from the
self-similarity depends on the definition of halo mass and
radius. Figure~\ref{fig:Tmw_nt_tot_overdensities} shows the normalized
mass-weighted, non-thermal, and total temperature profiles, $\Delta
\tilde{T} \equiv \tilde{T}(z)/\tilde{T}(z=0)-1$, for different
overdensity values: $\Delta = 1600, 500$, and $200$ defined with
respect to the mean density (left panels) and the critical density of
the universe (right panels). The range of the $x$-axis in each panel
shows the same physical radii corresponding to the radial range of
$0.1<R/R_{500c}<2.7$ at $z=0$.

The left panel of Figure~\ref{fig:Tmw_nt_tot_overdensities} shows that
the different choices of reference densities with respect to the mean
density results in varying degree of self-similar evolution in the
ICM. Specifically, we find that the departure from the self-similar
model diminishes for the smaller values of $\Delta_{\rm mean}$. For
example, the largest departure from the self-similar model is found
for the largest mean overdensity of $\Delta = 1600m$, where the
difference in the temperature at $R_{1600m}$ (which roughly
corresponds to $R_{500c}$ at $z=0$) is about $30\%$.  At $r/R_{\rm
  500m}=0.6$ in the second panel, the normalized temperature decreases
by $\sim 10\%$ from $z=0.7$ to $z=0$. At $r/R_{\rm 200m}=0.4$ in the
third panel, the evolution is even smaller (at the level of $\sim
5\%$).  Our results suggest that the overdensity of $\Delta=200m$
works best, because $R_{200m}$ tracks the average evolution of the
radial location of the accretion shock \citep{lau_etal15, shi16b},
where the bulk of thermalization process of the ICM occurs.

The right panel of Figure~\ref{fig:Tmw_nt_tot_overdensities}, on the
other hand, illustrates that the choice of different overdensities
with respect to the critical density does not significantly affect the
evolutionary trends in the inner regions, $r\lsim{R}_{500c}$, because
the critical density evolves slowly after $z\lsim1$ and tracks the
evolution of the gravitational potential in the inner regions.  The
gas beyond this radial range is less thermalized especially at high
redshifts due to the enhanced mass accretion, evidenced by the higher
normalization of the non-thermal temperature profiles (indicated by
dotted lines).  The total temperature (indicated by dashed lines), on
the other hand, remains self-similar with the difference of order
$10\%$ between $z=0.7$ and $z=0$.

\subsection{Effects of Substructures}
\label{sec:substructures2}

\begin{figure}
  \centering 
  \includegraphics[width=3.5in]{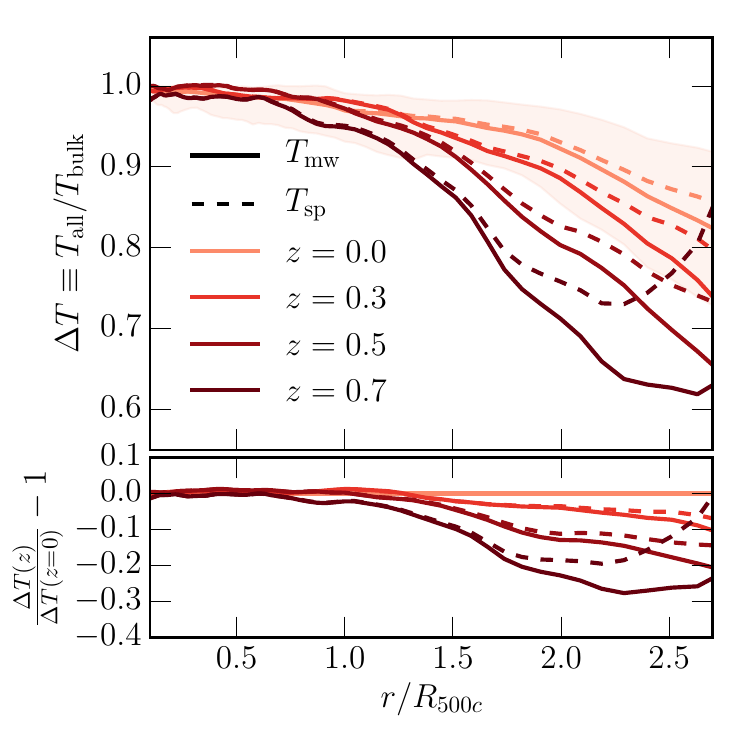}
  \vspace{-5mm}
  \caption{Top: Solid lines show the ratio of the mass-weighted
    temperatures $T_{\rm mw}$ of all gas to that of the substructure
    excluded ``bulk'' gas component at $z=0.0, 0.3, 0.5, 0.7$.  The
    dashed lines show the same ratio for the spectral-weighted
    temperature $T_{\rm sp}$. Bottom: Difference in the ratios of the
    temperature profiles with $z=0$.}\label{fig:ratios_bulk_mean}
\end{figure}

We quantify the effects of substructures on the temperature evolution
by comparing the average temperature profile, normalized to
$T_{500c}$, with and without excluding high density gas in infalling
subhalos and filaments.

The solid lines in Figure~\ref{fig:ratios_bulk_mean} correspond to the
ratios of $T_{\rm mw, all}$, the normalized mass-weighted temperature
of all gas, and $T_{\rm mw, bulk}$, the normalized mass-weighted
temperature of the substructure-excluded bulk component.  Each line
color corresponds to the average profile at $z=0.0, 0.3, 0.5$, and
$0.7$ for our simulated cluster sample.  The ratio $\Delta T_{\rm
  mw}=T_{\rm mw, all}/T_{\rm mw, bulk}$, is less than unity at all
radii and all redshifts because the dense gas associated with
substructures is typically cooler than the diffuse gas.  At
$r/R_{500c}=2$, $\Delta T_{\rm mw}$ is $92\%$ at $z=0$.  $\Delta
T_{\rm mw}$ decreases with redshift, reaching $70\%$ at $z=0.7$.

The dashed lines in the top panel of Figure~\ref{fig:ratios_bulk_mean}
show the ratio $\Delta T_{\rm spec}=T_{\rm spec, all}/T_{\rm spec,
  bulk}$, the spectral-weighted temperature ratios at $z=0.0, 0.3,
0.5$, and $0.7$.  Within $r/R_{500c}<1.5 $, $\Delta T_{\rm sp}$ are
nearly identical to those of $\Delta T_{\rm mw}$ at all redshifts. At
larger radii, $\Delta T_{\rm sp}$ is larger and evolves less than
$\Delta T_{\rm mw}$.  

The $\Delta T_{\rm sp}$ profiles indicate that spectral weighting
leads to the average of all of the gas to be more similar to the bulk
component.  Spectral weighting has both a temperature and density
dependence, whereas the mass weighting simply gives more weight to
cells that contain more gas mass.  The change in weighting places
slightly relatively more weight on the warmer bulk component of the
gas at large radii, decreasing the difference between $T_{\rm sp, all}$
and $T_{\rm sp, bulk}$.

In summary, substructures lead to a change in the average normalized
temperature profile at most $10\%$ at $r/R_{500c}=1.5$ from $z=0.7$ to
$z=0$, regardless of the weighting schemes.

\subsection{Effects of Non-Equilibrium Electrons}
\label{sec:ep}
\begin{figure}
  \centering 
  \mbox{
  \subfigure{\includegraphics[width=3.5in]{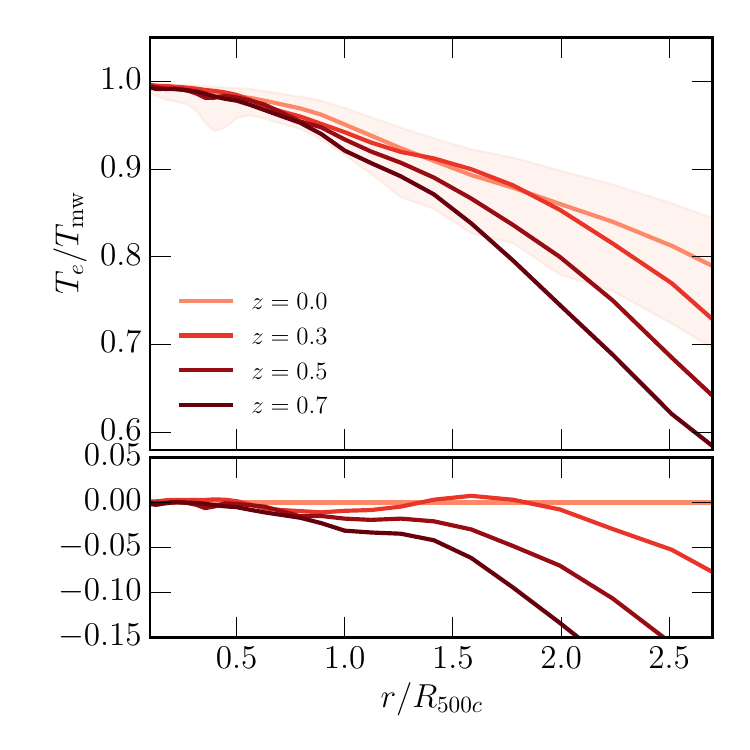}}
  }
  \vspace{-5mm}
  \caption{Top panel: Average ratio of the mass-weighted electron
    temperature, $T_e$, to the mass-weighted mean gas temperature,
    $T_{\rm mw}$, as a function of the cluster-centric radius at
    different redshifts: $z=0.0, 0.3, 0.5, 0.7$.  The shaded region
    indicates $1\sigma$ scatter in the $z=0$ profile.  Bottom panel:
    the fractional evolution in $T_e/T_{\rm mw}$ at high redshifts
    relative to $z=0$. }\label{fig:ep}
\end{figure}

The X-ray temperature is sensitive to the thermal energy of electrons
in the ICM plasma. However, since the equilibration time of electrons
and ions can be comparable to the Hubble time in the low-density
regions in the outskirts of galaxy clusters
\citep{spitzer_62,ruddandnagai_09}, the ICM temperature derived using
X-ray observation could be biased low.

The temperature bias from non-equilibrium electrons also depends on
the mass accretion rate and the mass of cluster
\citep{avestruz_etal15}.  On average, high redshift clusters have
higher mass accretion rates, which magnifies the bias.  However, high
redshift clusters have lower average masses.  The shorter Coulomb
equilibration timescale of the lower temperature gas in less massive
clusters therefore counteracts the effect of higher mass accretion
rates in high redshift clusters.  Figure~\ref{fig:ep} illustrates the
net effect, where we show profile of the ratio of the electron
temperature and the total gas temperature.  We thus conclude that the
evolution between $z=0$ and $z=0.7$ is less than $5\%$ up to
$r/R_{500c}\approx 1.5$.

\subsection{Effects of Baryonic Physics}
\label{sec:baryons}

\begin{figure}
  \centering 
  \mbox{
  \subfigure{\includegraphics[width=3.5in]{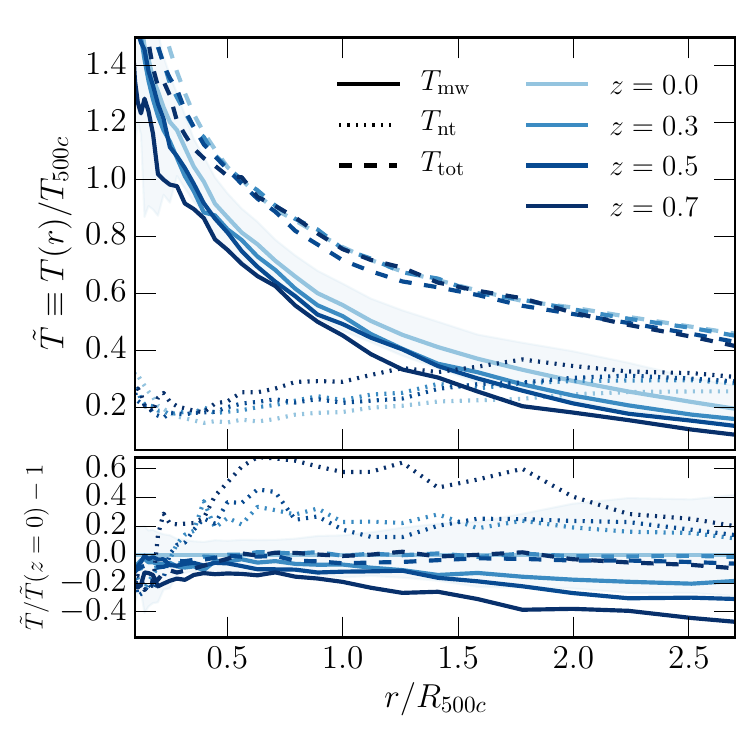}}
  }
  \vspace{-5mm}
  \caption{Similar to Figure~\ref{fig:Tmw_nt_tot}, showing the
    evolution of the temperature profiles for the simulated clusters
    with radiative cooling, star formation, and supernova feedback.  
    See the figure caption of Figure~\ref{fig:Tmw_nt_tot} for the 
    descriptions of panels and line types.
    }\label{fig:Tmw_nt_tot_CSF}
\end{figure}

Although the effects of baryonic heating and cooling processes are
expected to be small in the outskirts of galaxy clusters, dissipative
processes can potentially introduce additional physical scales that
can lead to breaking of self-similarity in the ICM properties.  To
assess these effects, we analyze a re-simulation of the {\em Omega
  500} box that includes radiative cooling, star formation, and
supernova feedback (CSF).  Since our CSF simulation does not include
feedback from active galactic nuclei, this simulation suffers from the
well-known ``overcooling'' problem.  Due to overcooling, too many
stars form in the cluster core, compared with observations, and the
simulation overestimates the impact of baryonic effects. The results
from our CSF simulations should therefore provide an {\em upper} limit
to the role of cooling and star formation.

Figure~\ref{fig:Tmw_nt_tot_CSF} shows the evolution of the ICM
temperature profiles in the CSF simulation. The normalized
mass-weighted temperature, $\tilde{T}_{\rm mw}$, at $r/R_{500c}\gtrsim
0.1$ evolves by about $20\%$ over the redshift range of $0\leq z \leq
0.7$, which is roughly consistent with the evolutionary trend seen in
the NR run. The non-thermal temperature profiles, $\tilde{T}_{\rm
  nt}$, in the CSF run (dotted lines in the upper panel of
Figure~\ref{fig:Tmw_nt_tot_CSF}) show a higher normalization in the
inner regions compared to their NR counterparts, due to rotational gas
motions induced by strong gas cooling in the CSF run \citep[see
  e.g.,][]{lau_etal11}. Despite the effects of baryonic physics on the
ICM temperature profiles, the total temperature profile in the CSF run
remains self-similar in the regions $0.2 \leq r/R_{500c}\leq 2.5$.
 
\begin{figure}
  \centering 
  \includegraphics[width=3.5in]{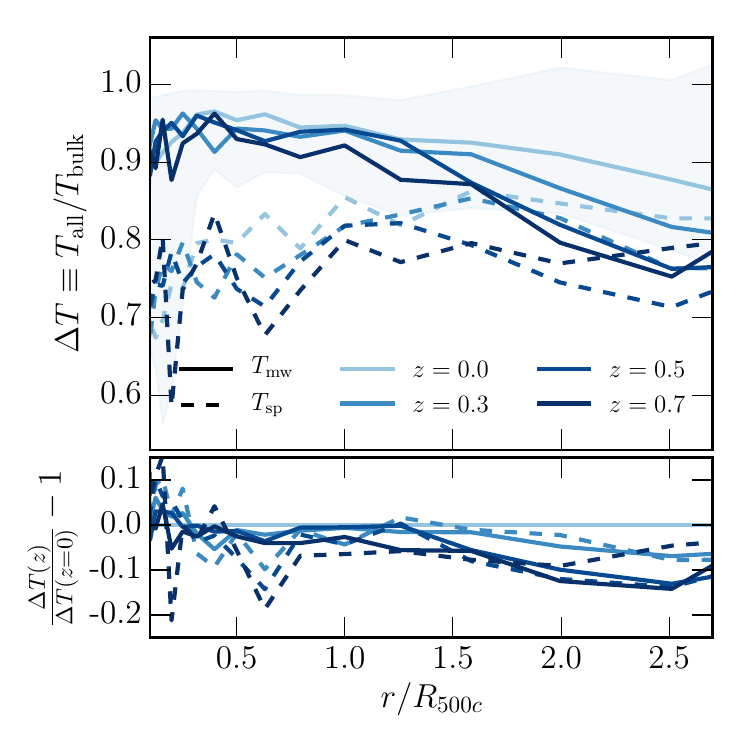}
  \vspace{-5mm}
  \caption{Same as Figure~\ref{fig:ratios_bulk_mean} for the CSF
    simulation.}\label{fig:ratios_bulk_mean_CSF}
  \vspace{5mm}
\end{figure}

Figure~\ref{fig:ratios_bulk_mean_CSF} shows the temperature ratio
$\Delta T\equiv {T}_{\rm all}/{T}_{\rm bulk}$ in the CSF simulation.
Similar to the results for the NR case in
Figure~\ref{fig:ratios_bulk_mean}, this ratio shows little evolution
for $r/R_{500c}\lesssim 1.5$ for both mass-weighted (solid lines) and
X-ray spectral-weighted temperatures (dashed lines). In fact, this
effect is smaller in the CSF run than in the NR run, with
only $\lesssim 15\%$ change between $0\leq z\leq 0.7$.

The profiles of both $\Delta T_{\rm mw}$ and $\Delta T_{\rm sp}$ have
lower normalizations in the CSF clusters than in the NR clusters,
since gas cooling in dense, gaseous substructures leads to lower
temperatures in the CSF run.  Unlike the NR profiles, $\Delta T_{\rm
  sp}$ at $r/R_{500c}\lesssim 1.5$ in the CSF run has a significantly
lower normalization than the profiles of $\Delta T_{\rm mw}$ at each
redshift, particularly in the inner radii.  Here, the spectroscopic
weighting gives more weights to substructures in the ICM that are
cooler and denser in the CSF run.  Nevertheless, the contribution of
substructures introduces little additional evolution in the ICM
temperature.

Thus, even in the presence of baryonic physics, the thermal
temperature evolution is mainly driven by the redshift-dependent
non-thermal gas motions, not by the increased amount of gaseous
substructures.

\section{Summary and Discussions}
\label{sec:conclusions}

We have investigated the origins of the evolution in the ICM
temperature profiles in the outskirts of galaxy clusters using a
mass-limited sample of 65 galaxy clusters extracted from the {\em
  Omega500} NR and CSF hydrodynamical cosmological simulations. Our
key findings are summarized below:

\begin{itemize}

\item The non-thermal pressure due to gas motions is the primary
  mechanism for driving the evolution of the ICM temperature profiles
  in the outskirts of galaxy clusters, producing a change in the ICM
  temperature by $30\%$ at $R_{500c}$ relative to the prediction of
  the self-similar model.

\item Gaseous substructures associated with infalling satellites and
  penetrating filaments contribute to the evolution of the temperature
  profiles by $\lesssim 10\%$ at $R_{500c}$, and it is subdominant to
  the evolution due to thermalization of gas motions.
  
\item The effects of spectral-weighted temperature, and the effects of
  non-equilibrium electrons, contribute to less than $10\%$ in the
  evolution of the ICM temperature.  Baryonic physics do not alter
  these conclusions.
    
\end{itemize}

These results suggest that the recently observed temperature evolution
in the outskirts of galaxy clusters by \citet{mcdonald_etal14} is
primarily due to the evolution of the non-thermal pressure profiles in
the ICM, in contrary to their original interpretation of
``superclumping''.

Our work further suggests that it may be possible to mitigate and
control the departure from the self-similar evolution.  First, we find
that an appropriate choice of cluster mass and radius can mitigate
departures from self-similarity in the temperature profile. The
reference density that scales with accretion in the outskirts, 200
times the mean background density, best captures self-similar
evolution in the temperature profile in the outskirts.  Other
reference densities with respect to the mean density, e.g.,
$\rho_{1600m}$ and $\rho_{500m}$, do {\em not} scale as well. Second,
we find that the ``total'' temperature, which is the sum of thermal
and non-thermal gas energies, exhibits a remarkable degree of
self-similar evolution when scaled with respect to the critical
density. 

In practice, while the thermal component of the ICM can be directly
imaged using X-ray and SZ observations, emerging high-angular
resolution X-ray and SZ imaging observations can measure fluctuations
in the ICM properties, which are sensitive to the level of non-thermal
pressure due to gas motions \citep[e.g.,][]{schuecker_etal04,
  khatri_gaspari16}. Future X-ray observatories, such as {\em
  Athena+}, also promise to provide direct measurements of gas motions
in the ICM through Doppler broadening of Fe lines.

Note that our simulations do not include other scale-dependent plasma
physics, such as thermal conduction and magnetic fields. However, we
expect these to have subdominant effects.  Thermal conduction in the
cluster outskirts is believed to be ineffective due to the long
conduction timescales of the diffuse low temperature outskirt gas
\citep[e.g.,][]{mccourt_etal13}.  While magnetic fields can drive gas
turbulence through magnetothermal instability (MTI)
\citep{parrish_etal12}, \citet{mccourt_etal13} shows MTI-driven
turbulence to be subdominant to the gas motions driven by mergers and
accretion for realistic cluster mass accretion histories. Therefore,
neither thermal conduction nor magnetic fields are likely to play
significant roles in the evolution of temperature profiles in cluster
outskirts.  Other effects, such as pressure provided by cosmic rays,
can in principle alter ICM properties in cluster outskirts.  We leave
the study of the effects of these plasma physics for future work.

To advance our understanding of ICM properties and their evolution in
cluster outskirts, future work should focus on (1) improving
theoretical modeling of both thermal and non-thermal components
including turbulence, cosmic rays, magnetic fields, and their
interactions; (2) deriving observational constraints on the
non-thermal temperature/pressure in the ICM based on pressure
fluctuations as well as direct measurements with the upcoming X-ray
missions; and (3) developing techniques to control the still poorly
understood astrophysical uncertainties and their impact on
cluster-based cosmological inferences.

\acknowledgments This work is supported by NSF grant AST-1412768, NASA
ATP grant NNX11AE07G, NASA Chandra Theory grant GO213004B, the
Research Corporation, and by the facilities and staff of the Yale
Center for Research Computing. CA acknowledges support from the
Fisk-Vanderbilt Dissertation Completion Award, the Kavli Institute of
Cosmological Physics, the Enrico Fermi Institute at the University of
Chicago, and the University of Chicago Provost's Office.
\lastpagefootnotes

\bibliography{ms}

\end{document}